\documentclass{itg}
\usepackage{amsfonts}
\usepackage{amsmath}
\usepackage{amssymb}
\usepackage{lscape}
\usepackage{epsf}
\usepackage{graphicx}

\newtheorem{theorem}{\indent Theorem}[section]

\newtheorem{proposition}[theorem]{\indent Proposition}

\newtheorem{EXAMPLE}{\indent Example}[section]
\newtheorem{definition}{\indent Definition}[section]

\newcommand{\code}{{\mathcal{C}}}
\newcommand{\graph}{{\mathcal{G}}}

\newcommand{\cV}{{\mathcal{V}}}
\newcommand{\cH}{{\mathcal{H}}}
 
\newcommand{\cE}{{\mathcal{E}}}

\newcommand{\cI}{{\mathcal{I}}}
\newcommand{\cJ}{{\mathcal{J}}}
\newcommand{\cN}{{\mathcal{N}}}
\newcommand{\cK}{{\mathcal{K}}}
\newcommand{\cP}{{\mathcal{P}}}
\newcommand{\cQ}{{\mathcal{Q}}}

\newcommand\rr{{\mathbb R}}

\newcommand{\bldc}{{\mbox{\boldmath $c$}}}
\newcommand{\bldcs}{{\mbox{\scriptsize \boldmath $c$}}}

\newcommand{\bldf}{{\mbox{\boldmath $f$}}}
\newcommand{\bldfs}{{\mbox{\scriptsize \boldmath $f$}}}
\newcommand{\bldh}{{\mbox{\boldmath $h$}}}

\newcommand{\bldp}{{\mbox{\boldmath $p$}}}

\newcommand{\bldS}{{\mbox{\boldmath $S$}}}
\newcommand{\bldss}{{\mbox{\scriptsize \boldmath $S$}}}

\newcommand{\bldT}{{\mbox{\boldmath $T$}}}
\newcommand{\bldP}{{\mbox{\boldmath $P$}}}
\newcommand{\bldpp}{{\mbox{\scriptsize \boldmath $P$}}}
\newcommand{\bldtt}{{\mbox{\scriptsize \boldmath $T$}}}

\newcommand{\bldw}{{\mbox{\boldmath $w$}}}
\newcommand{\bldx}{{\mbox{\boldmath $x$}}}
\newcommand{\bldX}{{\mbox{\boldmath $X$}}}
\newcommand{\bldy}{{\mbox{\boldmath $y$}}}
\newcommand{\bldz}{{\mbox{\boldmath $z$}}}

\newcommand{\zeros}{{\mbox{\boldmath $0$}}}

\newcommand{\rrr}{\mathfrak{R}}%
\newcommand{\rrrm}{\rrr^{-}}

\newcommand{\bldzero}{{\mbox{\boldmath $0$}}}

\newcommand{\Prob}{{p}}

\newcommand{\qed}{\hspace*{\fill}%
    \vbox{\hrule\hbox{\vrule\squarebox{.667em}\vrule}\hrule}\smallskip}
    \def\squarebox#1{\hbox to #1{\hfill\vbox to #1{\vfill}}}

\newlength{\Algwidth}

\renewcommand{\baselinestretch}{1.0}

\title{Linear-programming Decoding of Non-binary Linear Codes}
        
\author{Mark F. Flanagan$^1$, Vitaly Skachek$^2$,	Eimear Byrne$^3$, Marcus Greferath$^4$ \\[2mm]
      \mbox{}$^1$ Institute for Digital Communications, The University of Edinburgh, Edinburgh EH9 3JL, Scotland\\
      {\tt\small mark.flanagan@ieee.org} \vspace{2ex} \\
      \mbox{}$^{2,3,4}$
      Claude Shannon Institute, 
      University College Dublin, 
      Belfield,
      Dublin 4, Ireland \\
      \mbox{}$^2$ {\tt\small vitaly.skachek@ucd.ie} \\
      \mbox{}$^3$ {\tt\small ebyrne@ucd.ie} \\ 
      \mbox{}$^4$ {\tt\small marcus.greferath@ucd.ie}
}        

\begin{document}

\abstract{
We develop a framework for linear-programming (LP) decoding of non-binary linear codes over rings. We prove that the resulting LP decoder has the `maximum likelihood certificate' property, and we show that the decoder output is the lowest cost pseudocodeword. Equivalence between pseudocodewords of the linear program and pseudocodewords of graph covers is proved. LP decoding performance is illustrated for the $(11,6,5)$ ternary Golay code with ternary PSK modulation over AWGN, and in this case it is shown that the LP decoder performance is comparable to codeword-error-rate-optimum hard-decision based decoding.  
}

% \textbf{Keywords:} Pseudocodewords; Fundamental cone; Linear-programming decoder.

\maketitle

\section{Introduction}

For high-data-rate communication systems, bandwidth-efficient signalling schemes are required which necessitate the use of higher-order modulation. This may be achieved in conjunction with coding by the use of non-binary codes whose symbols map directly to modulation signals. 
A study of such codes over rings, particularly over the integers modulo $8$,
for use with PSK modulation was performed in~\cite{Deepak}. 

Of course, within such a framework it is desirable to use state-of-the-art error-correcting codes. 
\emph{Low-density parity-check} (LDPC) codes have become very popular in recent years due 
to their practical effectiveness under message-passing decoding. 
However, the analysis of LDPC codes is a difficult task. One approach was proposed in~\cite{Wiberg}, 
and it is based on the
consideration of so-called \emph{pseudocodewords} and their \emph{pseudoweights}. 
The approach was further explored in~\cite{FKKR},~\cite{KV-IEEE-IT}. In~\cite{Feldman-thesis} and~\cite{Feldman}, the
decoding of \emph{binary} LDPC codes using linear-programming decoding was proposed, and the connections 
between linear-programming decoding and classical belief propagation decoding were established. 
Recently, pseudocodewords of non-binary codes were defined and 
some bounds on the pseudoweights were derived in~\cite{Kelley-Sridhara-ISIT-2006}. 

In this work, we extend the approach in~\cite{Feldman} towards coded modulation, in particular to codes over rings mapped to non-binary modulation signals. As was done in~\cite{Feldman}, 
we show that the problem of decoding may be formulated as a linear-programming (LP) problem
for the non-binary case. 
We also show that an appropriate relaxation of the LP leads to a solution which has the `maximum likelihood (ML) certificate' property, i.e. if the LP outputs a codeword, then it must be the ML codeword. Moreover, we show that if the LP output is integral, then it must correspond to the ML codeword. We define the \emph{graph-cover pseudocodewords} of the code, and the \emph{LP pseudocodewords} of the code, and prove the equivalence of these two concepts. This shows that the links between LP decoding on the relaxed polytope and message-passing decoding on the Tanner graph generalize to the non-binary case. 

To demonstrate performance, LP decoding of the ternary Golay code is simulated, and the LP decoder is seen to perform approximately as well as codeword-error-rate optimum hard-decision decoding, and approximately $1.5$ dB from the union bound for codeword-error-rate optimum soft-decision decoding.

\section{General Settings}

We consider codes over finite rings (this includes codes over finite fields, but may be more general). 
Denote by $\rrr$ a ring with $q$ elements, by $0$ its additive identity, and let $\rrrm = \rrr \backslash \{ 0 \}$.     
Let $\code$ be a linear $[n,k]$ code with parity-check matrix $\cH$ over~$\rrr$. 
The parity check matrix $\cH$ has $m \ge n-k$ rows.
 
Denote the set of column indices and the set of row indices of $\cH$ by  $\cI = \{1, 2, \cdots, n \}$ 
and $\cJ = \{1, 2, \cdots, m \}$, respectively. 
We use notation $\cH_j$ for the $j$-th row of $\cH$. 
Let the graph $\graph = (\cV, \cE)$ 
be the Tanner graph of $\code$ associated with the matrix $\cH$, namely $\cV = \{u_1, u_2, \cdots, u_n \} \cup  
\{v_1, v_2, \cdots, v_m \}$, and there is an edge between $u_i$ and $v_j$ if and only if $\cH_{j,i} \neq 0$. 
We denote by $\cN(v_j)$ the set of neighbors of the vertex $v_j$, and by $\mbox{supp}(\bldc)$ the support of 
a vector $\bldc$. 
%For a word $\bldx \in \rrr^n$, we use a notation $(\bldx)_S$ for a subword indexed by 
%the set of indices $S \subseteq \cI$. 
Let $d = \max_{j \in \cJ} \{ |\mbox{supp}(\cH_j)| \}$. 

For a word $\bldc  = (c_1, c_2, \cdots, c_n) \in \rrr^n$, 
we associate the value $c_i$ with variable vertex $u_i$ for each $i \in \cI$. 
Parity-check $j\in\cJ$ is said to be 
\emph{satisfied} if and only if $\sum_{i\in\cI} \cH_{j,i} \cdot c_i = 0$. 
We say that the vector $\bldc$ is a codeword 
of the single parity-check code $\code_j$ if and only if parity check $j \in \cJ$ is satisfied.
Also, we say that the vector $\bldc$ is a codeword 
of $\code$ if and only if all parity checks $j\in\cJ$ are satisfied.  

\begin{definition}
(\cite{KV-characterization})
A graph $\tilde{\graph} = (\tilde{\cV}, \tilde{\cE})$ is a \emph{finite cover} of the graph $\graph = (\cV, \cE)$ 
if there exists a mapping $\Pi: \tilde{\cV} \rightarrow \cV$ which is a graph homomorphism
($\Pi$ takes adjacent vertices of $\tilde{\graph}$ to adjacent vertices of $\graph$), such that 
for every vertex $v \in \graph$ and every $\tilde{v} \in \Pi^{-1}(v)$, the neighborhood $\cN(\tilde{v})$ 
of $\tilde{v}$ is mapped bijectively to $\cN(v)$. 
\end{definition}

\begin{definition}
(\cite{KV-characterization})
A cover of the graph $\graph$ is called an $M$-cover, where $M$ is a positive integer, if $|\Pi^{-1}(v)| = M$
for every vertex $v \in \cV$.  
\end{definition}
 
Fix some positive integer $M$. Let $\tilde{\graph} = (\tilde{\cV}, \tilde{\cE})$ be an $M$-cover   
of the graph $\graph = (\cV, \cE)$ representing 
the code $\code$ with parity-check matrix $\cH$. 
Denote the vertices in the sets $\Pi^{-1} (u_i)$ and $\Pi^{-1} (v_j)$ by 
$\{ u_{i,1}, u_{i,2}, \cdots, u_{i,M} \}$ and $\{ v_{j,1}, v_{j,2}, \cdots, v_{j,M} \}$, respectively, where $i\in\cI$ and $j\in\cJ$.

Consider the linear code $\tilde{\code}$ of length $Mn$ over $\rrr$, 
defined by the $Mm \times Mn$ parity-check matrix $\tilde{\cH}$. For 
$1 \le i^*,j^* \le M$ and $i \in \cI$, $j \in \cJ$, we let  
$i' = (i-1) M + i^*, j' = (j-1) M + j^*$, and
\[
\tilde{\cH}_{j',i'} = \left\{ \begin{array}{cl}
\cH_{j,i} & \mbox{if } u_{i,i^*} \in \cN(v_{j,j^*}) \\
0 & \mbox{otherwise} 
\end{array} \right. \; .
\]
Then, any vector $\bldp \in \tilde{\code}$ has the form 
\begin{eqnarray*}
 \bldp & = & ( p_{1,1}, p_{1,2}, \cdots, p_{1,M}, p_{2,1}, p_{2,2}, \\
 && \hspace{3ex} \cdots, p_{2,M}, \cdots, p_{n,1}, p_{n,2}, \cdots, p_{n, M} ) \; .
\end{eqnarray*}
We associate the value 
$p_{i,\ell} \in \rrr$ with the vertex $u_{i,\ell}$ in $\tilde{\graph}$ ($i\in\cI$, $\ell = 1, 2, \cdots, M$).

The word $\bldp \in \tilde{\code}$ as above is called a \emph{graph-cover pseudocodeword} 
of the code $\code$. 
Sometimes, we consider the following $n \times q$ 
matrix representation, denoted $\cP$, of the pseudocodeword $\bldp$: 
\[
\Big( m_i (\alpha) \Big)_{i \in \cI; \, \alpha \in \rrr} \; ,  
\] 
where 
\[
m_i(\alpha) = \left| \{ \ell \in \{ 1, 2, \cdots, M \} \; : \; p_{i,\ell} = \alpha \} \right| \ge 0 \; , 
\]
for $i\in\cI$, $\alpha \in \rrr$.

\section{Decoding as a Linear-Programming Problem} 
 
Assume throughout that the codeword $\bar{\bldc} = (\bar{c}_1, \bar{c}_2, \cdots, \bar{c}_n) 
\in \code$ has been transmitted over a $q$-ary input
memoryless channel,
and a corrupted word $\bldy = (y_1, y_2, \cdots, y_n) \in \Sigma^n$ has been received. Here $\Sigma$ denotes the set of channel output symbols; we assume that this set either has finite cardinality, or is equal to $\mathbb{R}^l$ or $\mathbb{C}^l$ for some integer $l \ge 1$. In practice, this channel may represent the combination of modulator and physical channel.
We assume hereafter that all information words are equally probable, and so all codewords are 
transmitted with equal probability. 

It was suggested in~\cite{Feldman-thesis} to represent each symbol as a binary vector of 
length $|\rrrm|$, where the entries in the vector are indicators of a symbol 
taking on a particular value. Below, we elaborate on this approach.  
It should be mentioned that by using such a representation, the non-binary code is converted into a binary code. 
However, this binary code is not linear, and therefore the analysis in~\cite{Feldman-thesis},~\cite{Feldman}
is not directly applicable. 

For use in the following derivation, we shall define the mapping 
\[
\xi \; : \; \rrr \longrightarrow \{ 0, 1 \}^{q-1} \subset \mathbb{R}^{q-1} \; , 
\]
defined by 
\[
\xi (b) = \bldx = ( x^{(\alpha)} )_{\alpha \in \rrrm } \; , 
\]
such that, for all $\alpha \in \rrrm$,
\[
x^{(\alpha)}=\left\{ \begin{array}{cc}
1 & \textrm{ if } b = \alpha \\
0 & \textrm{ otherwise }\end{array}\right. \; . 
\]
We note that the mapping $\xi(\cdot)$ is one-to-one, 
and its image is the set of binary vectors of length $q-1$ with Hamming weight 0 or 1.

We also define a function $\boldsymbol{\lambda}: \Sigma \longrightarrow {\mathbb R} \cup \{\pm\infty\}$ by
\[
\boldsymbol{\lambda} = ( \lambda^{(\alpha)} )_{\alpha \in \rrrm}  \; , 
\]
where, for each $y \in \Sigma$, $\alpha \in \rrrm$,
\[
\lambda^{(\alpha)}(y) = \log \left( \frac{\Prob ( y | 0 )}{ \Prob ( y| \alpha ) } \right) \; , 
\]
and $p(y|c)$ denotes the channel output probability (density) conditioned on the channel input. 
Extend $\boldsymbol{\lambda}$ to a map on $\Sigma^n$ 
by $\boldsymbol{\lambda}(\boldsymbol{y})= (\boldsymbol{\lambda}(y_1) \;|\; \boldsymbol{\lambda}(y_2) \;|\; 
\ldots \;|\; \boldsymbol{\lambda}(y_n))$.

The codeword-error-rate-optimum receiver operates according to the \emph{maximum a posteriori} (MAP) decision rule:
\begin{eqnarray*}
\hat{\bldc} & = & \arg \max_{\bldcs \in \code} \Prob ( \; \bldc \; | \; \bldy \; ) \\
 & = & \arg \max_{\bldcs \in \code} \frac{\Prob ( \; \bldy \; | \; \bldc \; ) \Prob ( \; \bldc \; )}{\Prob ( \; \bldy \; )} \; .
\end{eqnarray*}
Here $\Prob\left(\cdot\right)$ denotes probability if $\Sigma$ has finite cardinality, and probability density if $\Sigma$ has infinite cardinality.

By assumption, the \emph{a priori} probability $\Prob ( \bldc )$ is uniform over codewords, and $\Prob ( \bldy )$ is independent of $\bldc$. Therefore, the decision rule reduces to maximum likelihood (ML) decoding:
\begin{eqnarray*}
 \hat{\bldc} & = & \arg \max_{\bldcs \in \code} \Prob ( \; \bldy \; | \; \bldc \; ) \\
 & = & \arg \max_{\bldcs \in \code} \prod_{i=1}^n \Prob ( y_i | c_i ) \\
 & = & \arg \max_{\bldcs \in \code} \sum_{i=1}^n \log ( \Prob ( y_i | c_i ) )
\end{eqnarray*}
\begin{eqnarray*}
\phantom{\hat{\bldc}}
 & = & \arg \min_{\bldcs \in \code} \sum_{i=1}^n \log \left( \frac{\Prob ( y_i | 0 )}{\Prob ( y_i | c_i )} \right) \\
 & = & \arg \min_{\bldcs \in \code} \sum_{i=1}^n \boldsymbol{\lambda}(y_i) \xi (c_i) ^T \; ,
\end{eqnarray*}
where we have made use of the memoryless property of the channel, and of the fact that if 
$c_i = \alpha \in \rrrm$, then $\boldsymbol{\lambda}(y_i) \xi (c_i)^T = \lambda^{(\alpha)}(y_i)$. This is then equivalent to
\begin{equation}
\begin{split} 
( \xi(\hat{c}_1) \; | \; \xi(\hat{c}_2)  \; | \;  \ldots & \; | \; \xi(\hat{c}_n) ) \\
& = \arg \min_{\bldfs \in \cK(\code)} \sum_{i=1}^n \boldsymbol{\lambda}(y_i) \bldf_i^T \\
& = \arg \min_{\bldfs \in \cK(\code)} \boldsymbol{\lambda}(\boldsymbol{y}) \bldf^T \; , 
\label{eq:object-function}
\end{split}
\end{equation}
where
$$
\bldf = ( \bldf_1 \; | \; \bldf_2 \; | \; \cdots \; | \; \bldf_n )$$ and $$\bldf_i = ( f_i^{(\alpha)} )_{\alpha \in \rrrm} \; \mbox{for all $i \in \cI$} \; , 
$$
and where  $\cK(\code)$ represents the convex hull of all points $\bldf\in\rr^{(q-1)n}$ which correspond to codewords, i.e.
\[ 
\cK(\code) = H_\mathrm{conv} \big\{ (\xi(c_1) \; | \; \xi(c_2) \; | \; \ldots \; | \; \xi(c_n)) \; : 
\; \bldc\in\code \big\} \; . 
\]
Therefore it is seen that the ML decoding problem reduces to the minimization of a linear objective function (or cost function) over a polytope in $\rr^{(q-1)n}$.
The number of variables and constraints for this linear program is exponential in $n$, and it is therefore too complex for practical implementation. 
To circumvent this problem, we formulate a relaxed LP problem, as shown next. 

The solution we seek for $\bldf$ (i.e. the desired LP output) is
\[
\bldf = ( \xi(\bar{c}_1) \; | \; \xi(\bar{c}_2)  \; | \; \ldots  \; | \; \xi(\bar{c}_n) ) \; .
\]

We introduce auxiliary variables whose constraints, along with those of the elements of $\bldf$,
will form the relaxed LP problem. 
First, for each $j \in \cJ$, we define the mapping $\bldX_j(\bldc)$ of the words $\bldc \in \rrr^n$, 
$\bldX_j(\bldc) = ( X_{j,\alpha}(\bldc))_{\alpha \in \rrrm}$, where 
\[
X_{j,\alpha} (\bldc) = \{ i \in \mbox{supp}( \cH_j ) \; : \; c_i = \alpha \} \; , 
\]
for $\alpha \in \rrrm$. 
For each word $\bldc \in \rrr^n$, $X_{j,\alpha}(\bldc)$ is the set of word indices 
where symbol $\alpha$ appears in parity check $j$, for $j \in \cJ$, $\alpha \in \rrrm$.  
We define the set $E_j$ as 
\begin{eqnarray*}
E_j & = & \{ \bldS = (S_\alpha)_{\alpha \in \rrrm} = \bldX_j (\bldc) \; : \; \bldc \in \code_j \big\} \; . 
\end{eqnarray*}
In other words, $\bldX_j(\bldc) \in E_j$ if and only if parity
check $j$ is satisfied by the word $\bldc \in \rrr^n$. 

We now introduce the auxiliary variables 
\[
w_{j,\bldss} \; \mbox{ for } \;  j\in\cJ, \bldS \in E_j \; ,
\]
and denote the vector containing these variables as
\[
\bldw = \big( \; w_{j,\bldss} \; \big)_{j \in \cJ, \bldss \in E_j} \; ,  
\]
with respect to some ordering on the elements of $E_j$. 
The solution we seek for these variables is 
\begin{eqnarray*}
\forall j \in \cJ & : & w_{j,\bldss} = 
\left\{ \begin{array}{cl}
1 & \textrm{if } \bldS = \bldX_j(\bar{\bldc}) \\
0 & \textrm{otherwise }\end{array}\right. . 
\end{eqnarray*}
To this end, we impose the constraints
\begin{eqnarray}
\forall j \in \cJ, \; \forall \bldS \in E_j,  \quad  0 \le w_{j,\bldss} \le 1 \; ,
\label{eq:equation-polytope-3} 
\end{eqnarray} 
and
\begin{equation}
\forall j \in \cJ, \quad \sum_{\bldss \in E_j} w_{j,\bldss} = 1 \; .
\label{eq:equation-polytope-4} 
\end{equation} 

Finally, we note that the solution we seek satisfies the further constraints 
\begin{eqnarray}
&  \forall j \in \cJ, \; \forall i \in \mbox{supp}(\cH_j), \; \forall \alpha \in \rrrm, \nonumber \\
& f_i^{(\alpha)} = 
\sum_{\bldss \in E_j, \; i \in S_\alpha} w_{j,\bldss} \; .
\label{eq:equation-polytope-5} 
\end{eqnarray} 

Constraints~(\ref{eq:equation-polytope-3})-(\ref{eq:equation-polytope-5}) form a polytope which we denote $\cQ$. The minimization of
the objective function~(\ref{eq:object-function}) over $\cQ$ forms the relaxed LP decoding problem.
This LP is defined by $O(qn + q^d m)$ variables and $O(qn + q^d m)$ constraints. 
We note that the further constraints 
\begin{equation}
\forall i \in \cI, \; \forall \alpha \in \rrrm, \quad 0 \le f_i^{(\alpha)} \le 1 \; ,
\label{eq:equation-polytope-1} 
\end{equation} 
and
\begin{equation}
\forall i \in \cI, \quad \sum_{\alpha \in \rrrm} f_i^{(\alpha)} \le 1 \; ,
\label{eq:equation-polytope-2} 
\end{equation} 
follow from the constraints~(\ref{eq:equation-polytope-3})-(\ref{eq:equation-polytope-5}), for any 
$(\bldf, \bldw) \in \cQ$.

Now we may define the decoding algorithm, which works as follows. 
The decoder solves the LP problem of minimizing the objective 
function~(\ref{eq:object-function}) subject to the constraints~(\ref{eq:equation-polytope-3})-(\ref{eq:equation-polytope-5}).  
If $\bldf \in \{0,1\}^{(q-1)n}$, the output is the codeword $(\xi^{-1}(\bldf_1), \xi^{-1}(\bldf_2), \cdots , \xi^{-1}(\bldf_n) )$ (we shall prove in the next section that 
this output is indeed a codeword).
Otherwise, the decoder outputs an `error'. 

%-------------------------------------------------------------------------------------- 

\section{Polytope Properties} 

The analysis in this section is a direct generalization of the results in~\cite{Feldman}. 

\begin{definition}
An \emph{integral point} in a polytope is a point with all \emph{integer} coordinates.  
\end{definition}

\begin{proposition}
$\,$
\begin{itemize} 
\item[1)]
Let $(\bldf, \bldw) \in \cQ$, and $f_i^{(\alpha)} \in \{ 0,1 \}$ for every $i \in \cI$, $\alpha \in \rrrm$. 
Then, 
\[
( \xi^{-1} ( \bldf_1 ) \; , \xi^{-1} ( \bldf_2 ) \; , \cdots \; , \xi^{-1} ( \bldf_n ) ) \in \code \; .
\]
\item[2)]
Conversely, for every codeword $\bldc = (c_1, c_2, \cdots, c_n) \in \code$, there exists 
$\bldw$ such that $(\bldf, \bldw)$ is an integral point in $\cQ$ with 
$\bldf_i = \xi(c_i)$ for all $i \in \cI$.
\end{itemize}
\end{proposition} 
 
{\bf Proof.} 
\begin{enumerate}
\item
Suppose $(\bldf, \bldw) \in \cQ$, and 
$f_i^{(\alpha)} \in \{ 0,1 \}$ for every $i \in \cI$, $\alpha \in \rrrm$.

Define $\bldc$ by $c_i = \xi^{-1}(\bldf_i)$ for all $i \in
\cI$. By~(\ref{eq:equation-polytope-2}), this is well defined. 
Define $\bldT = (T_\alpha)_{\alpha \in \rrrm} = \bldX_j(\bldc)$, i.e.  
\begin{equation}
T_\alpha = \{ i \in \mbox{supp}(\cH_j) \; : \; f^{(\alpha)}_i = 1 \} \; , 
\label{eq:T_k_sets_definition} 
\end{equation}
for $\alpha \in \rrrm$. Now, fix some $j\in\cJ$ and let $\bldP = (P_\alpha)_{\alpha \in \rrrm} \in E_j$, $\bldP \neq \bldT$. There must exist $\alpha \in \rrrm$
and $i_0 \in \cI$ such that either $i_0 \in P_\alpha \backslash
T_\alpha$ or $i_0 \in T_\alpha \backslash P_\alpha$. 

If $i_0 \in P_\alpha \backslash
T_\alpha$, then by~(\ref{eq:equation-polytope-5}) and~(\ref{eq:T_k_sets_definition})
\[
f^{(\alpha)}_{i_0} = 0 = \sum_{\bldss \in E_j , \; i_0 \in S_\alpha} w_{j, \bldss} \; .
\]
Therefore $w_{j,\bldss} = 0$ for all $\bldS\in E_j$ with $i_0 \in S_\alpha$,
and in particular $w_{j,\bldpp} = 0$. 

If $i_0 \in T_\alpha \backslash
P_\alpha$, then
by~(\ref{eq:equation-polytope-4}),~(\ref{eq:equation-polytope-5}), and~(\ref{eq:T_k_sets_definition})
\begin{eqnarray*}
0 & = & 1 - f_{i_0}^{(\alpha)} \\
& = & \sum_{\bldss \in E_j} w_{j,\bldss} -  
\sum_{\bldss \in E_j, \; i_0 \in S_\alpha} w_{j,\bldss} \\
& = & \sum_{\bldss \in E_j, \; i_0 \notin S_\alpha} w_{j,\bldss} \; .
\end{eqnarray*}
Therefore $w_{j,\bldss} = 0$ for all $\bldS \in E_j$ with $i_0 \notin S_\alpha$,
and in particular $w_{j,\bldpp} = 0$. 

It follows that $w_{j,\bldss} = 0$ for all $\bldS \in E_j$, $\bldS \neq
\bldT$. But by~(\ref{eq:equation-polytope-4}) this implies that $\bldT
\in E_j$ (and that $w_{j,\bldtt} = 1$). Applying this argument for every $j\in\cJ$ implies
$\bldc \in \code$.
\item
For $\bldc \in \code$, we let $\bldf_i = \xi(c_i)$ for $i \in
\cI$. For each parity check $j\in\cJ$, we let $\bldT = (T_\alpha)_{\alpha \in \rrrm} = 
\bldX_j(\bldc) \in E_j$ and then set
\begin{eqnarray*}
\forall j \in \cJ : && w_{j,\bldss} = 
\left\{ \begin{array}{cl}
1 & \textrm{if } \bldS = \bldT \\
0 & \textrm{otherwise. }\end{array}\right. 
\end{eqnarray*}
It is easily checked that the resulting point $(\bldf, \bldw)$ is integral and satisfies
constraints~(\ref{eq:equation-polytope-3})-(\ref{eq:equation-polytope-5}).
\qed
\end{enumerate}

The following proposition assures the so-called \emph{ML certificate} property. 
\begin{proposition}
Suppose that the decoder outputs a codeword $\bldc \in \code$. Then, $\bldc$ 
is the maximum-likelihood codeword. 
\end{proposition} 

The proof of this proposition is straightforward. The reader can refer to a similar proof for 
the binary case in~\cite{Feldman}. 

%-------------------------------------------------------------------------------------- 
 
\section{Transmission-Independent Decoder Performance} 
In this section, we state a theorem on decoder performance, namely, that under a certain symmetry condition, the probability of decoder failure is independent of the transmitted codeword. Decoder failure is defined as the event where the decoder output is not equal to the transmitted codeword (this could correspond to a non-integral value of $\bldf$, or to an erroneous output codeword).

{\bf Symmetry Condition.}  

For each $\alpha\in\rrr$, there exists a bijection 
\[
\tau_{\alpha} \; : \; \Sigma \longrightarrow \Sigma \; , 
\]
such that the channel output probability (density) conditioned on the channel input satisfies
\begin{equation*}
p(y|\beta) = p(\tau_{\alpha}(y)|\beta-\alpha) \; ,
%\label{eq:symmetry_condition}
\end{equation*}
For all $ y\in \Sigma$, $\beta \in \rrr$.
When $\Sigma$ is equal to $\mathbb{R}^l$ or $\mathbb{C}^l$ for $l \ge 1$, the mapping $\tau_{\alpha}$ is assumed to be isometric with respect to Euclidean distance in $\Sigma$, for every $\alpha\in\rrr$.
\begin{theorem}
Under the stated symmetry condition, the probability of decoder failure is independent of the transmitted codeword.
\label{htrm:equl-prob} 
\end{theorem}
The proof of this theorem is omitted due to space limitations. Examples of modulator-channel combinations for which this assumption holds are: $q$-ary PSK modulation over AWGN (where the additive group of $\rrr$ is cyclic); orthogonal modulation over AWGN; and the discrete memoryless $q$-ary symmetric channel. 

\section{Linear-Programming Pseudo-codewords} 
\begin{definition}
A \emph{linear-programming pseudo-codeword} (LP pseudocodeword) of the code $\code$ is a vector $(\bldh, \bldz)$ where 
\[
\bldh = ( \bldh_1 \; | \; \bldh_2 \; | \; \cdots \; | \; \bldh_n ) \; , 
\]
\[
\forall i \in \cI , \; \bldh_i = ( h_i(\alpha) )_{\alpha \in \rrrm} \; , 
\]
\[
\bldz = \big( \; z_{j,\bldss} \; \big)_{j \in \cJ, \bldss \in E_j} \; ,  
\]
\end{definition}
where the elements of $\bldz$ are nonnegative integers, and the following two conditions hold for all $j \in \cJ$:
\begin{eqnarray}
& \forall i \in \mbox{supp}(\cH_j), \; \forall \alpha \in \rrrm, \hspace{8ex} \nonumber \\
& h_i(\alpha) = \sum_{\bldss \in E_j, \; i \in S_\alpha} z_{j,\bldss} \; ,
\label{eq:equation-polytope-new} 
\end{eqnarray} 
\begin{eqnarray}
\forall i \in \mbox{supp}(\cH_j), \;  
h_i(0) = \sum_{\stackrel{\bldss \in E_j}{\forall \alpha \in \rrrm \; : \; i \notin S_\alpha}} 
z_{j,\bldss} \; .
\label{eq:equation-polytope-new-2} 
\end{eqnarray} 
From~(\ref{eq:equation-polytope-new}) and~(\ref{eq:equation-polytope-new-2}) it follows that the elements of $\bldh$ are nonnegative integers, and that for each $i \in \mbox{supp}(\cH_j) \cap \mbox{supp}(\cH_{j'})$, we have
\begin{equation}
\sum_{\alpha \in \rrr} h_i(\alpha) = \sum_{\bldss \in E_j} z_{j,\bldss} = \sum_{\bldss \in E_{j'}} z_{j',\bldss} \; . 
\label{eq:equation-constant-sum-h} 
\end{equation}
We assume that the Tanner graph of $\cH$ is connected; it then follows 
from~(\ref{eq:equation-constant-sum-h}) that 
\[
\forall i \in \cI \; : \; \sum_{\alpha \in \rrr} h_i(\alpha) = M \; , 
\]
for some fixed nonnegative integer $M$. 

We note that the LP pseudocodeword $(\bldh, \bldz)$ defined above can be represented by 
the $n \times q$ matrix 
\[
\mathsf{H} = \Big( h_i (\alpha) \Big)_{i \in \cI; \, \alpha \in \rrr}  \; .
\] 

In the following, we say that the decoder \emph{fails} if the decoder output is not equal to the transmitted codeword.
\begin{theorem}
Assume that the all-zero codeword was transmitted. 
\begin{enumerate}
\item
If the LP decoder fails, then there exists some LP
pseudocodeword $(\bldh, \bldz)$, $\bldh \neq \zeros$, such that 
\begin{equation}
\sum_{i=1}^n \left( \sum_{\alpha \in \rrrm}  \lambda^{(\alpha)}(y_i) h_i(\alpha) \right) \le 0 \;. 
\label{eq:optimal-pseudo}
\end{equation} 
\item
If there exists some LP pseudocodeword $(\bldh, \bldz)$, $\bldh \neq \zeros$, such that 
\begin{equation}
\sum_{i=1}^n \left( \sum_{\alpha \in \rrrm} \lambda^{(\alpha)}(y_i) h_i(\alpha) \right) < 0 \; ,
\label{eq:optimal-pseudo-small}
\end{equation} 

then the LP decoder fails.  
\end{enumerate}
\end{theorem}

{\bf Proof.} The proof follows the lines of its counterpart in~\cite{Feldman}. 
\begin{enumerate}
\item
Let $(\bldf, \bldw)$ be the point in $\cQ$ which minimizes $\boldsymbol{\lambda}(\boldsymbol{y}) \bldf^T$. Suppose the decoder fails; then $\bldf \neq \bldzero$, and we must have   
$\boldsymbol{\lambda}(\boldsymbol{y}) \bldf^T \le 0$.

%Then,   
%\begin{equation}
%\sum_{i=1}^n \left( \sum_{k=1}^{q-1}  \lambda_i^{(k)} f_i^{(k)} \right) \le 0 \; . 
%\label{eq:optimal-pseudo-2}
%\end{equation} 

Next, we construct the LP pseudocodeword $(\bldh, \bldz)$ as follows. Since the LP has rational coefficients, all elements of the vectors $\bldf$ and $\bldw$ must be rational. Let $M$ denote their lowest common denominator; since $\bldf \neq \bldzero$ we may have $M>0$. Now set $h_i (\alpha) = M \cdot f_i^{(\alpha)}$ for all $i \in \cI$, $\alpha \in \rrrm$, 
set $z_{j,\bldss} = M  \cdot w_{j,\bldss}$ for all $j \in \cJ$ and $\bldS \in E_j$, and then define $h_i (0)$ 
as in~(\ref{eq:equation-polytope-new-2}) for all $i \in \cI$.
By~(\ref{eq:equation-polytope-3}) and~(\ref{eq:equation-polytope-5}), $(\bldh, \bldz)$ is an LP pseudocodeword and $\bldh \neq \bldzero$ since $\bldf \neq \bldzero$. Also $\boldsymbol{\lambda}(\boldsymbol{y}) \bldf^T \le 0$ implies~(\ref{eq:optimal-pseudo}).
\item
Now, suppose that an LP pseudocodeword $(\bldh, \bldz)$ with $\bldh \neq \bldzero$ satisfies~(\ref{eq:optimal-pseudo-small}). 
Let 
\[
M = \sum_{\alpha \in \rrr} h_i(\alpha) \; . 
\]
Since $\bldh \neq \bldzero$ we have $M>0$. Now:
\begin{itemize}
\item 
Set $f_i^{(\alpha)} = h_i(\alpha) / M $ for all $i \in \cI$, $\alpha \in \rrrm$;  
\item 
Set
$w_{j,\bldss} = z_{j,\bldss} / M$  
for all $j \in \cJ$ and $\bldS \in E_j$. 
\end{itemize}
It is straightforward to check that $(\bldf, \bldw)$ satisfies all the constraints of the polytope 
$\cQ$. Also, $\bldh \neq \zeros$ implies $\bldf \neq \zeros$. 
Finally, ~(\ref{eq:optimal-pseudo-small}) implies $\boldsymbol{\lambda}(\boldsymbol{y}) \bldf^T < 0$. Therefore, the LP decoder will produce an output other than the all-zero codeword, resulting in decoder failure. 
\end{enumerate}
\qed

\section{Equivalence Between Pseudo-codeword Sets}

In this section, we show the equivalence between the set of LP pseudocodewords and the set of
graph-cover pseudocodewords. The result is summarized in the following theorem. 

\begin{theorem}
There exists an LP pseudocodeword $(\bldh, \bldz)$ for the code $\code$ with 
matrix representation $\mathsf{H}$ if and only if there 
exists a graph-cover pseudocodeword $\bldp$ with the same matrix representation. 
\end{theorem}

{\bf Proof.} 
\begin{enumerate}
\item
Let $(\bldh, \bldz)$ be an LP pseudocodeword, and   
let $\graph = (\cV, \cE)$ be the Tanner graph $\code$ associated with the parity-check matrix $\cH$. 
We define 
\[
M = \sum_{\alpha \in \rrr} h_i(\alpha) \; . 
\]
(Recall that under our assumption that the Tanner graph is connected, the value of $M$ is independent of $i$.)
Below, we construct a corresponding $M$-cover graph $\tilde{\graph} = (\tilde{\cV}, \tilde{\cE})$. 

\begin{itemize}
\item
For every $i \in \cI$, and for every $\alpha \in \rrr$, 
the graph $\tilde{\graph}$ will contain $h_i(\alpha)$ copies 
of the vertex $u_i$ associated with the value $\alpha$. 
\item
For every $j \in \cJ$, $\bldS \in E_j$, 
the graph $\tilde{\graph}$
will contain $z_{j,\bldss}$ copies of the check vertex $v_j$, associated with 
the $(q-1)$-tuple $\bldS$. 
\item
The edges in the graph are connected according to the membership in the sets $S_\alpha$, for $\alpha \in \rrrm$. 
Namely, each copy of check vertex $v_j$ will be connected to one copy of $u_i$ for every $u_i \in \cN(v_j)$. 
A copy of a check vertex $v_j$ associated with the $(q-1)$-tuple $\bldS$
will be connected to a copy of $u_i$ associated with the value $\alpha \in \rrrm$ if and only if $i \in S_\alpha$.
A copy of $v_j$ associated with the $(q-1)$-tuple $\bldS$
will be connected to a copy of $u_i$ associated with the value $0$ if and only if 
$i \notin \cup_{\alpha \in \rrrm} S_\alpha$. 
\end{itemize}

By using~(\ref{eq:equation-polytope-new}), we see that for every $j \in \cJ$, $i \in \mbox{supp}(\cH_j)$, 
$\alpha \in \rrr$, 
there are exactly $h_i(\alpha)$ edges 
connecting the copies of the vertex $v_j$ with the copies of $u_i$ associated with the value $\alpha$. 
Therefore, the graph $\tilde{\graph}$ is well-defined, and the neighborhood of a copy of $v_j$
contains exactly one copy of $u_i$ for every $u_i \in \cN(v_j)$. 
Furthermore, it can be seen that the neighborhood of a copy of $u_i$
contains exactly one copy of $v_j$ for every $v_j \in \cN(u_i)$.
In addition, all copies of all check vertices $v_j$ represent satisfied checks, and  
therefore $\bldp$, induced by the graph $\tilde{\graph}$,
is a graph cover pseudocodeword of $\code$, as claimed. 

\item
Now let $\bldp$ be a graph-cover pseudocodeword corresponding to some 
$M$-cover of the Tanner graph of $\code$. Then, 
\begin{itemize}
\item
for every $i \in \cI$, and for every $\alpha \in \rrr$, 
we define $h_i(\alpha)$ to be the number of copies of the vertex $u_i$ associated with value $\alpha$.
\item
for every $j \in \cJ$, and for every $\bldS \in E_j$,
we define $z_{j,\bldss}$ to be the number of copies of the check vertex $v_j$ 
connected to copies of $u_i$, associated with $\alpha \in \rrrm$ for $i \in S_\alpha$, 
and associated with $0$ for $i \notin \cup_{\alpha \in \rrrm} S_\alpha$. 
\end{itemize}

Then, $z_{j,\bldss}$ are all nonnegative integers 
for all $j \in \cJ$ and $\bldS \in E_j$. Moreover,~(\ref{eq:equation-polytope-new}) and~(\ref{eq:equation-polytope-new-2}) hold for all $j \in \cJ$ by construction of the graph. Therefore, $(\bldh, \bldz)$ is an LP pseudocodeword of the code $\code$.  
\end{enumerate}
\qed

\section{Simulation Study}

\begin{figure}[hbt]
\begin{center}\includegraphics[%
  width=1.0\columnwidth,
  keepaspectratio]{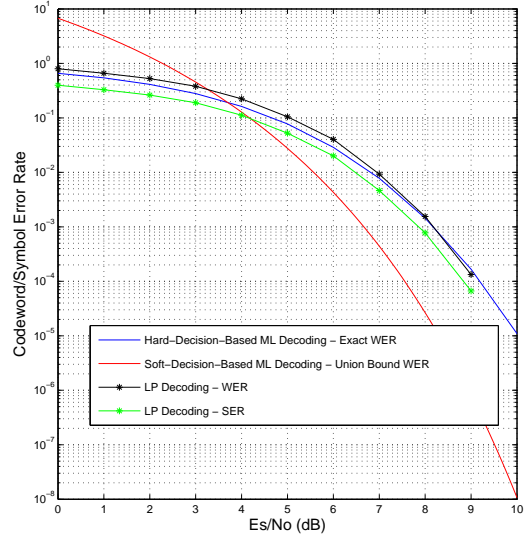}
\end{center}

\caption{\label{cap:Golay}Codeword error rate (WER) and symbol error rate (SER) for the $\left(11,6,5\right)$ ternary
Golay code under $3$-PSK modulation. The figure shows performance
under LP decoding, as well as the exact result for hard-decision decoding
and the union bound for soft-decision decoding.}
\end{figure}

In this section we compare performance of the linear-programming decoder with hard-decision and soft-decision based ML decoding. For such a comparison, a code and modulation scheme are needed which possess sufficient symmetry properties to enable derivation of analytical ML performance results. We consider encoding of $6$-symbol blocks
according to the $\left(11,6,5\right)$ ternary Golay code, and modulation
of the resulting ternary symbols with $3$-PSK modulation prior to
transmission over the AWGN channel. The symbol error rate (SER) and
codeword error rate (WER) are shown in Figure \ref{cap:Golay}. To
quantify performance, we define the signal-to-noise ratio (SNR) per
information symbol $\gamma_{s}=E_{s}/N_{0}$ as the ratio of receive
signal energy per information symbol to the noise power spectral density.
Also shown in the figure are two other performance curves for WER.
The first is the exact result for ML hard-decision decoding of the
ternary Golay code; since the Golay code is perfect, this is obtained
from 
\[
\textrm{WER}(\gamma_s)=\sum_{\ell=3}^{11}\binom{11}{\ell}(p(\gamma_s))^{\ell}\left(1-p(\gamma_s)\right)^{11-\ell} \; , 
\]
where $p(\gamma_s)$ represents the probability of incorrect hard decision at
the demodulator and was evaluated for each value of $\gamma_s$ using numerical
integration. The second WER curve represents the union bound for ML
soft-decision decoding. Using the symmetry of the $3$-PSK constellation,
this may be obtained from 
\[
\textrm{WER}(\gamma_s)<\frac{1}{2}\sum_{\bldcs \in \code}
\textrm{erfc }\left(\sqrt{\frac{3}{4}w_{H}(\bldc) \; r\gamma_{s}}\right) \; , 
\]
where $r$ denotes the code rate, and the Hamming weight of the codeword $\bldc \in \code$, $w_{H}(\bldc)$, 
is given by the weight enumerating polynomial 
\[
W\left(x\right)=1+132x^{5}+132x^{6}+330x^{8}+110x^{9}+24x^{11} \; . 
\]
The performance of LP decoding is approximately the same as that of
codeword-error-rate optimum hard-decision decoding. The performance
lies $0.1$ dB from the result for ML hard-decision decoding and
$1.53$ dB from the union bound for codeword-error-rate optimum soft-decision
decoding at a WER of $10^{-4}$.

\section*{Acknowledgements} 

This work was supported by the Claude
Shannon Institute for Discrete Mathematics, Coding and Cryptography (Science Foundation Ireland Grant 06/MI/006). The authors would like to thank J.~Feldman and R.~Koetter for helpful discussions.

%-----------------------------------------------------------------------

%-----------------------------------------------------------------------

\end{document}